# Measurement of Cloud Properties Using a Self-Designed Cloud Chamber

*Ridhesh Goti*[1]*, *Bhashin Thakore*[1], and *Rohit Srivastava*[2]

[1]Department of Physics, Pandit Deendayal Petroleum University, Gandhinagar, 382007, Gujarat, India.
[2]Faculty, Department of Science, Pandit Deendayal Petroleum University, Gandhinagar, 382007, Gujarat, India.

**Abstract.** Present paper examines the dependency of ambient parameters such as humidity, and turbulence to determine the conditions on raindrop formation with the help of a self-designed cloud chamber. The research methods are experimental and observational in nature, where atmospheric phenomena are recreated through the usage of appropriate substitutes. Miniature droplets were created inside a box-like setup through the use of dry ice to cool the water vapor rising up, so as to create suspended water droplets, and to induce precipitation of heavier droplets. The experiment resulted in the creation of precipitated droplets, which were found at the base of the chamber at 99-100% relative humidity. The suspended droplets were used to study factors such as luminosity and variation of droplet sizes with turbulence. It was found that up to 14.4 m/s of turbulence, the droplet sizes increase with an increase in turbulence, with the luminosity decreasing with increase in turbulence. The gaussian profile of droplet size distribution has also been obtained, with a standard deviation of 2.83, 3.01 and 3.18 for low, medium and high turbulence speeds respectively. The experiment can be extended to incorporate a higher number of variables, so as to include a wider range of atmospheric phenomena. **Keywords.** Cloud chamber, Droplet measurement, Turbulence, Luminosity, Droplet distribution

**Acknowledgements**
**We are thankful to Mr. Rutul Patel who provided great expertise and guidance in the technical and manufacturing process of the cloud chamber. We also place our thanks for the generous assistance provided by Mr. Dhaval Santola and Hitansh Shah, without whom the experiment could not have been conducted. We are thankful to Ms. Ruchita Shah for providing invaluable insights to improve the paper.**

## 1 Introduction

Clouds are essential in the earth-atmosphere system as they regulate earth's radiation budget by reflecting as well as by scattering the solar radiation and they also absorb infrared radiation of the Earth. They are required for precipitation to occur and hence are essential part in the hydrological cycle. For cloud to form, they require hygroscopic nuclei for condensation through water vapor, which then triggers drop formation process in the atmosphere. Due to mixing of different types of aerosols, the cloud formation processes vary as aerosol act as a seed to the cloud. Reflection of solar radiation by the cloud depends upon the amount of liquid water and concentration of liquid water droplets present in the cloud. Other indirect processes involve cloud cover in UV scattering and the role of gases such as sulphur oxide and nitrogen oxide in light absorption and attenuation was elaborated upon by Spinhirne and Green in their 1978 publication [1]. Using satellite observations, the overall role of clouds in radiative fluxes at the top of the atmosphere, at the surface, and at levels within the atmosphere and climate was explained in a comprehensive report by Bruce A Wielicki in 1995 [2]. Characteristics of the cloud are highly dependent on various parameters such as types of aerosol, mixing of different types of aerosol, humidity level, concentration of cloud condensation nuclei, size-concentration of cloud droplets and cloud turbulence. Cloud formation is a huge complex process and are highly dependent on seasons. It is quite important to understand the stability of all these processes to avoid climatic catastrophes.

Study of these processes can be done using satellite data, model simulation and *in*-situ measurements. Among them, satellite data gets restrict with temporal resolution and ground-based data limits with spatial resolution. To avoid such circumstances, model simulation study plays a significant role as all these processes are space and time dependent. Thus, the present study focuses on studying these processes using a self- designed cloud chamber, which helps to simulate various realistic scenarios. The cloud chamber may give better results, which may improve our understanding about cloud and rain formation. Thus, the study of cloud formation processes in the cloud chamber under given conditions may play a significant role to understand cloud properties in the atmosphere.

The study of cloud formation starts with the first cloud chamber which was originally designed by C.T.R Wilson in 1911 [3] to observe and photograph particle tracks, their potential in the field of cloud physics was undeniable. Further the study on understanding the nucleation of gaseous molecules was demonstrated by Katz (1970) [4]. Later on, study of scattering and reflection properties of hollow ice particles in Manchester Ice Cloud Chamber was carried out by Helen Smith et al. [5]. With time, researchers have reported the turbulent exchange of flow in the cloud chamber which was observed by Cholemari and Arakari (2005) [6]. Due to turbulence, cloud and precipitation processes, earth's radiation budget also varies which in turn would affect the temperature of the Earth [7]. From the study by intergovernmental panel on climate change (IPCC) in 2013 [8], the role of clouds in radiation budget of the Earth is variable. In these variations, water droplets play the central role in albedo phenomena due to clouds. Experimental data from (European Cloud and Radiation Experiment) EUCREX [9] reported that size distribution of water droplets also have their impact on cloud albedo. Like precipitation, albedo also depends on the physical as well as chemical processes of aerosol which were studied by Chang et al. in 2017 [10].

The cloud chambers used to simulate atmospheric processes have evolved through the years. The transition of the cloud chamber from a detector of particle tracks to a simulator of atmospheric processes started with the expansion cloud chamber [11]. Subsequently, a variety of chambers like aerosol with a continuous flow system ([12], [13]) , turbulence-based chambers ([14]) and wind tunnels ([15]) were invented through modifications in the expansion chamber. The main purpose of these cloud chambers was to study various cloud properties using high calibration instruments for measuring atmospheric parameters such as effect of turbulence, presence of aerosols in the chamber, and other essential processes such as cloud formation and precipitation.

A new way of precipitation through He-Ne laser has been used for condensation of water droplets in the cloud chamber and is further opted for the present study Rohwettter et al. (2010) [16], and Petit et al. (2011) [17]. Both ideas are based on temperature gradient inside the chamber. This method differs from silver iodide cloud seeding technique in the atmosphere B. J. Mason in 1975 [18]. Such cloud chambers have capability to generate physical simulations of the numerous phenomena exist in the atmosphere. By varying the factors such as humidity, light attenuation through aerosol concentration and temperature of the region might affect the formation of clouds, which in turn lead to affect weather as well as climate patterns of the region. Such cloud chambers have been instrumental in weather prediction and analysis numerous times, and have been used time and again to simulate conditions physically, so as to obtain a better idea about the concomitant results as opposed to computer simulations. It remained to be seen, however, if the processes could be created on a smaller scale, and if it could display phenomena similar to that of the larger cloud chamber. There have also been attempts by several agencies ([12-15], [19], [20]) to map such atmospheric processes. Therefore, in an attempt to simulate these atmospheric processes in a different manner, a cloud chamber was designed for the explicit purpose of recreating these processes and recording observations based on the processes incorporated into the cloud chamber.

The aim of the experiment is to observe the changes in temperature with the height, and to describe the droplet formation and distribution with changes in the height and turbulence. The experiment also aims to measure the attenuation of light with changes in turbulence. As the turbulence increases, the change in the intensity of the light should decrease due to higher droplet separation. A sufficient number of droplets should yield a gaussian distribution, with droplets of certain diameters being more prominent in number than the other droplets. This has been elaborated upon in the subsequent sections. The results from the experiment satisfy the aforementioned hypotheses.

## 2 Experimental Setup

The study of this paper is focused on the formation of small cloud droplets inside the chamber of size 0.55 m X 0.6 m X 1 m, and studying properties with its help. The chamber is made up of acrylic sheets with a thickness of 5 mm. In most of the cold clouds, the nucleation process begins at temperatures below -40 °C inside low-level clouds. However, the process here carries out the cloud formation procedures with a minimum temperature of -8 °C, which is commonly found in mid-level clouds like the alto cumulus and alto status cloud types. These types of clouds contain ice particles and water droplets. The temperature to -8 °C was achieved with the help of dry ice and a fan built inside a separate cooling box. Here, the fan is used to introduce varying levels of turbulence inside the chamber to simulate mid-level clouds to an acceptable accuracy. The fan has three control values to regulate its speed. Usually, the upper part of the cloud is at a relatively higher temperature compared to the lower part, so the cooling system is arranged at the base of the chamber.

The water vapor is introduced into the chamber at a height of 0.8 m from base. The water vapor is injected through a plastic pipe with an airtight valve from the boiler. The boiler in question heats up the water at around 70 °C to generate the vapor. A hygrometer with a resolution 0.1 °C and range -40 °C to 100 °C was used for measuring the temperature with humidity. The hygrometer measured the variables from top to bottom of the chamber. Varying the height and fan speed allows the study of droplet formation at different levels of humidity and turbulence. Also, to measure luminosity of laser light with wavelength of 632.8 nm, a Lux meter has been used. The droplet sizes were captured and analysed using a DSLR camera of a resolution of 5184 X 3456 or 18 megapixels. The primary objective was to obtain clouds or miniature sized droplets using the schematic design given in Fig. 1.

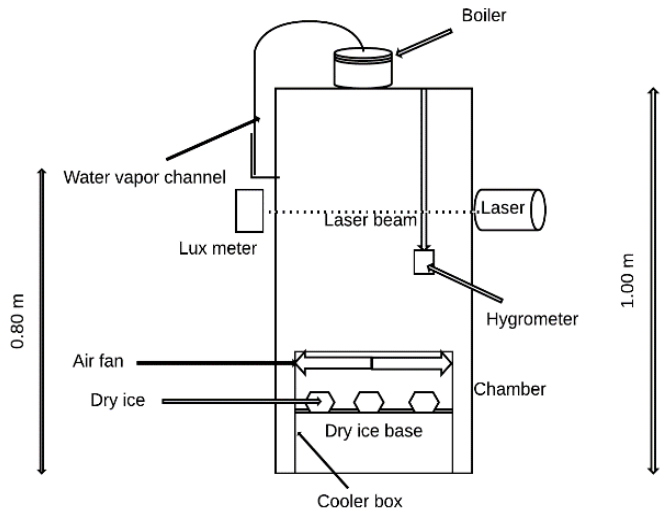

**Fig. 1.** A schematic of the self-designed cloud chamber.

## 3 Procedure

For setup of experiment, the cooler box was firstly adjusted and enclosed within chamber. The pipe inlet was joined with boiler and hygrometer was passed through middle of the chamber. Oil was applied on one side of the chamber to avoid fogging, and to observer clear distinction in the droplets that accumulated on the surface. The stages of the experiment are elaborated as given below.

The first stage of the experiment consists of introducing water vapour inside the cloud chamber. This was achieved by passing steam through a pipe. The water vapour acts as a substitute for the suspended moisture in the atmosphere. The temperature and humidity when it entered the chamber were measured using a hygrometer.

The second stage involved the introduction of turbulence inside the cloud chamber. A fan inserted inside the chamber was used to achieve the same, with three different speeds to gauge the turbulence. This led to the droplets colliding with each other, and in turn, undergoing coagulation through collision.

The third stage involved the cooling of the cloud chamber. Dry ice was used to introduce temperature gradients of appreciable magnitudes inside the chamber. This led to the cooldown of the suspended droplets, which in turn would lead to precipitation.

The fourth stage was the observational stage. A laser was shone through a cavity in the cloud chamber, and the luminosity was measured on the opposite side. This led to the discovery of possible light attenuation. For the experiment, we have used light of 632.8nm, incident from a Helium-Neon laser. The droplets resultant due to the precipitation were photographed for further analysis, and the post precipitation humidity was measured.

The experiment was conducted for three different cases, each involving a different turbulence speed. In the first part, the fan was configured to have a relatively low speed of 2.80 m/s. In the first part, water vapor from the externally positioned boiler was passed through a pipe and injected into the cloud chamber from the top. As mentioned before, the temperature of the water vapor was set to 70 °C. Because of the dry ice at the bottom, the vapor condensed and small droplets were found to have accumulated at the bottom of the chamber. For the sake of accuracy, instead of measuring the relatively mercurial droplet behaviour inside the chamber, the camera was used to photograph droplet accumulation on the surface of the chamber. Humidity and temperature were measured through the hygrometer with variation of height inside the chamber. The luminosity was measured using a lux meter, with the source of light being a Helium-Neon laser of a wavelength of 632.8 nm. The second and third parts of the experiment involved similar procedures, but with turbulence speeds of 7.20 m/s and 14.40 m/s, respectively.

.

## 4 Observations and Results

Through the experiments conducted, the observations of humidity, temperature, lux intensity of laser light, and droplet sizes were measured with various instruments and software. These results were analysed with different turbulent environment and heights. The droplet size distribution, temperature gradient was plotted from the analysis. The readings were taken after cooling the chamber for a time of 7 minutes (measured using a stopwatch), wherein the temperature inside the chamber would stabilize, making the chamber suitable for measurement and analysis. Using a digital thermometer, it was determined that the initial temperature was 18 °C. Next, with the help of dry ice and the fan, a minimum temperature of -7 °C was achieved within 9.5 minutes. The system was allowed to stabilize for seven minutes after that, following which the bottom of the chamber was at a temperature of -6 °C. The variations were observed that while the temperature did not vary significantly with different fan speeds, it did change noticeably with height. Also, the initial humidity of the chamber was 50 % and it rose to 99 % (as measured using a digital hygrometer) at the outlet of water vapor after 7 min. Owing to the instrumental limitations of the hygrometer, the humidity could not be measured beyond 99% [21]. Therefore, it is assumed that the relative humidity crosses 100% at the time of precipitation [21]. The results of each step were measured after 4 minutes of stabilization time, and were tabulated as shown.

The first three table take into account the changes in relative humidity inside the cloud chamber due to the injection of water vapor along with variation in low, medium and high fan speeds. Table 1, for example, holds the data recorded for the temperature and relative humidity at a turbulence speed of 2.8m/s while Tables 2 and 3 display data for the same quantities (albeit with turbulence speeds of 7.2 m/s and 14.4 m/s respectively).

**Table 1.** Variation of temperature and humidity with height at a turbulence of 2.8 m/s

| Height (cm) | Temperature (°C) | Relative humidity (%) |
|---|---|---|
| 0 | -6.15 | 96 |
| 20 | -0.15 | 97 |
| 40 | 8.35 | 99 |
| 60 | 16.40 | 99 |
| 80 | 23.00 | 99 |

**Table 2.** Variation of temperature and humidity with height at a turbulence of 7.2 m/s

| Height (cm) | Temperature (°C) | Relative humidity (%) |
|---|---|---|
| 0 | -6.15 | 96 |
| 20 | -0.8 | 97 |
| 40 | 7.50 | 99 |
| 60 | 15.3 | 99 |
| 80 | 22.2 | 99 |

**Table 3.** Variation of temperature and humidity with height at a turbulence of 14.4 m/s

| Height (cm) | Temperature (°C) | Relative humidity (%) |
|---|---|---|
| 0 | -6.15 | 95 |
| 20 | -1.20 | 98 |
| 40 | 7.00 | 99 |
| 60 | 14.80 | 99 |
| 80 | 20.85 | 99 |

It was noticed that the variation in humidity is nearly zero after 0 cm, owing to the fact that the relative humidity inside a cloud must be 100%. It is important to note that the device used to measure the humidity had an upper limit of 99%, which means that the humidity may have been well over 100%. This can be researched further in similarly constructed cloud chambers. Fig. 2, as obtained from a measurement using digital thermometer, graphically demonstrates a relationship between the height and temperature in low, medium and high turbulence speeds. It is interesting to note that the relationship was found to be nearly linear in all three cases. At the high turbulence, gradient of temperature from top to bottom is decreased. The changes in luminosity with an increase in turbulence speeds were observed, as shown in Table 4, using a lux meter. It can be seen that the scattering is highest at low turbulences. These results of light attenuation are matched with the reference data [22].

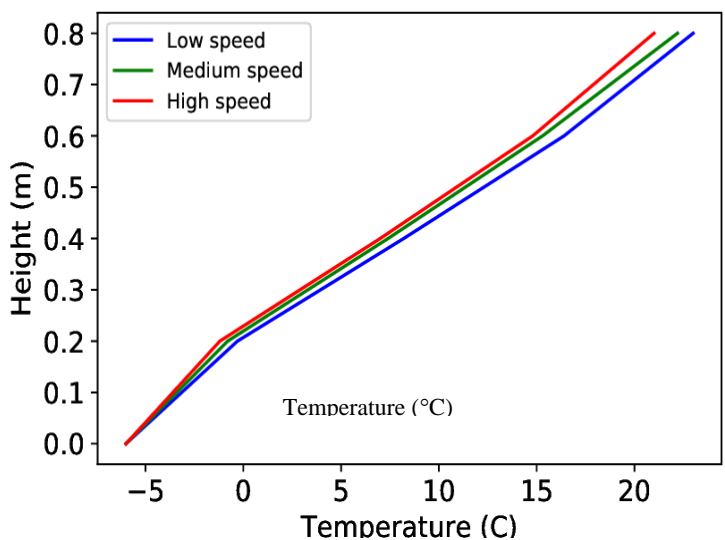

**Fig. 2.** Graphical representation of the variation of temperature with change in height in low, medium and high turbulence speeds.

**Table 4.** Variation of luminosity with turbulence

| Turbulence Speeds(m/s) | Observed luminosity at 0.80 m (lux) | Percentage change (%) |
|---|---|---|
| 2.8 | 70 | 76.67 |
| 7.2 | 103 | 65.67 |
| 14.4 | 125 | 58.34 |

It is important to know how the sizes of the droplets vary with turbulence, so as to determine a possible relationship, and in the process, develop a better understanding of how clouds form. The distribution of droplet inside the chamber is very important factor. The distribution for all three cases were got using DSLR camera and software, by analysing the size of one droplet in the picture, and using the obtained length to count other droplets in the obtained distribution. The droplet diameters for all three cases were counted using a software called 'ImageJ' [23], [24], and the distribution was obtained using data points in Microsoft Excel [25].

Fig. 3 displays the variation in droplet sizes with fan turbulence. It can be observed that the droplets that formed at the highest fan speeds had the highest average size, at a diameter of 4-4.243 mm, while the droplets that formed at low fan speeds, that is, at lower turbulence, had a relatively lower average size, at approximately 3.162-3.464 mm. In the Fig. 3, the results are drawn between frequency and droplet size in mm. The analogy of these distributions is plotted using gaussian curve or equation. The gaussian-esque profile has been achieved by putting appropriate values of mean value and standard deviation.

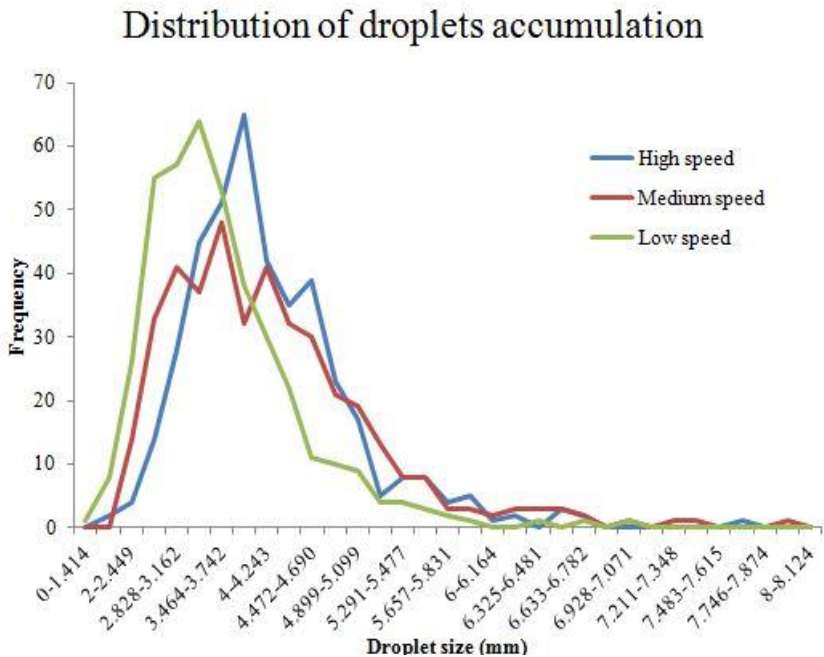

**Fig. 3.** Variation of droplet sizes with turbulence

In the Fig. 4, the distribution is between frequency and the surface areas of each droplet, obtained using a method similar to the one used above. Therefore, Fig. 4 draws a parallel between the empirically obtained distribution and the ideal gaussian distribution. The standard deviation calculated for low, medium and high turbulence speeds were found to be

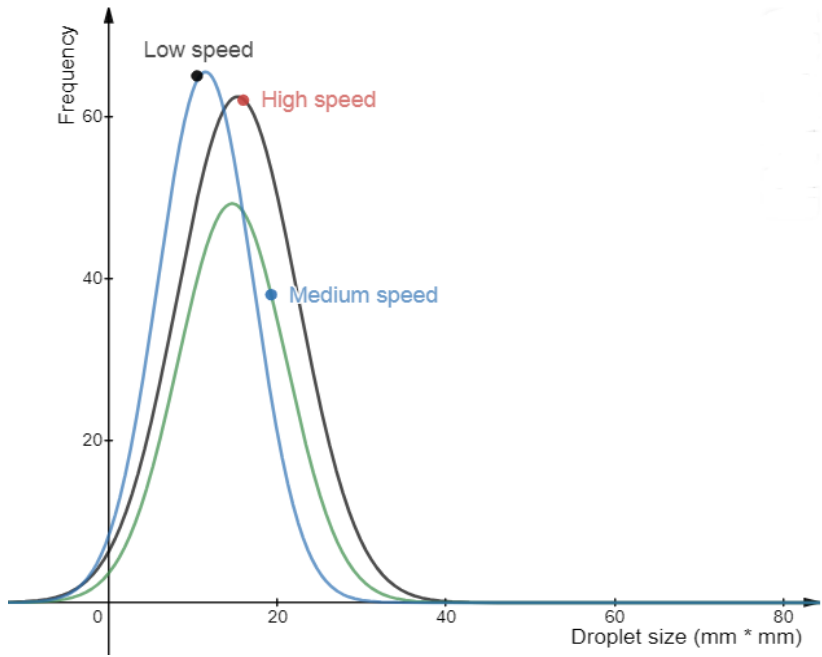

**Fig. 4.** Visualizing the Gaussian-curve properties of the droplet distribution at low, medium and high turbulence speeds. (Here x-axis is area of the droplets in mm2)

2.83, 3.01 and 3.188mm respectively. The droplet distributions obtained in the chamber are displayed in Fig. 5. The parameters are given in Table 5. Here, $N$ is the scaling factor for gaussian curve to match with our results. It is taken as the number of particles per unit volume. From the observation, the mean radius and standard deviation increases with turbulence or factors which are equivalent to turbulence.

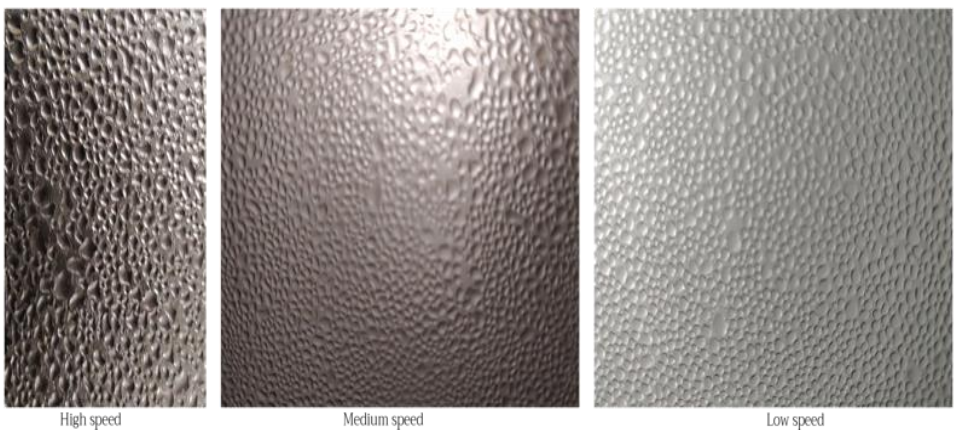

**Fig. 5.** Droplet distributions obtained for High, Medium and Low Speeds

Table 5. Parameters of the Gaussian Profile

| Speed | N | Mean radius (mm) | Standard deviation (mm) |
|---|---|---|---|
| Low | 1310 | 3.39 | 2.83 |
| Medium | 1120 | 3.83 | 3.01 |
| High | 1580 | 3.91 | 3.18 |

The results of the size distribution and the effect of the turbulence are reconciled with reference data in [26], [27]. After Gaussian fits, the effect of the turbulence on standard deviation and mean value greatly matched with [26], [27]. The Gaussian fit was done in open accessible graphical calculator [28].

## 5 Conclusion

The experiment and the designing of the cloud chamber was made with the purpose to provide a different approach to the construction of a miniaturized cloud chamber, so as to replicate basic atmospheric processes. From the tables and the figures obtained, it can be concluded sufficiently that the results have been achieved to a certain degree of accuracy, especially when we keep in mind the relatively simple design of the cloud chamber. The effect of turbulence on the temperature gradient is analysed in Fig. 2. This indicates that with changing the gradient by turbulence, the cloud's dynamics are altered, mainly, parameters such as light attenuation and droplet size distribution are affected. The observed change in the luminosity of the incident light due to scattering is given in the Table IV. This percentage change, however, is also dependent on wavelength, and light rays with smaller wavelengths will experience higher scattering, thus displaying a greater difference between the luminosity of the measured light, and the luminosity of the incident light. This variation is minimum when turbulence is 14.4 m/s, however, it does not indicate that the source should get progressively luminous with turbulence as the speed is increased beyond 14.4 m/s. This concludes that turbulence makes the atmosphere in such dynamic that the probability of the scattering of light reduced because of the random motion of the system. Also, the droplet distribution graph tells us about the maximum size of particles one can obtain when there is a specific amount of turbulence. The largest droplet sizes were obtained when turbulence at 14.4 m/s. This is believed to happen because of an increased rate of collision-coalescence properties between two droplets. For this process too, there may be a threshold of turbulence speeds over which the droplet sizes begin to reduce, and disperse due to extreme, unfavourable turbulence speeds. Overall, the idea of the turbulence through air fan and cooling by dry ice makes the design very easy to simulate the atmospheric changes due to turbulence. Also, very few studies were found on light attenuation due to atmosphere, specifically for simulation in cloud chambers.

## 6 Discussion and Future Scope

In the present study, the setup was restricted to a simple cloud chamber for basic simulation of small droplet accumulation and to study the effects of turbulence on droplets distributions along with other parameters like temperature and humidity. In the experiment, the simple way to generate the clouds is water vapor pressure and the temperature gradient. Using this concept and some instruments, the results are obtained. The main variable parameter is the turbulence and it is defined in very simple way in this work. The mathematics of the turbulence such as the naiver stokes field equations have not been included. But the basic involvement of the turbulence through air fan provided the expected results, which have been shown to match results from similar previous studies [22], [26], [27]. The results showed that changes in luminosity of laser light was 76.67 %, 65.67 %, and 58.34 % at low, medium, and high speeds respectively. This indicated that with higher turbulence, the changes in the luminosity of light are reduced. The results show great matching of the trend with reference data. Also, the temperature gradient when transitioning between the upper and lower parts of the chamber is 29 °C, 28.2 °C, 27 °C at low, medium, and high speed, respectively. The gradient of temperature also increases with decreasing fan speeds. This displays a property of the cooling effects of the dry ice "scattering" at the lower levels as the turbulence is increased. Meanwhile, the temperature at the upper part of the cloud chamber decreases with fan speed. The changes of these parameters are unidirectional with respect to a constant fan speed. Also, the droplet size distribution for the different cases of turbulence intensity have been analysed. Many techniques are used for the droplet size distribution such as the use of a Disdrometer, but here, a high-resolution camera is used to capture the miniature droplets inside the chamber. The taken pictures for different turbulent intensity, the size of the droplets and their concentration are analysed through scientific tools. The Gaussian fit with of these results gives the exact idea on effects of turbulence on the distribution's mean value and standard deviation. There are many mathematical models regarding the droplet size distribution for the clouds. Many distributions like log-normal and Gaussian are obtained during the observations. These distributions depend on the cloud's temperature. Based on that, the fitting of the Gaussian profiles for the different turbulent intensity are successfully created. The atmosphere inside the chamber is inhomogeneous and contains aerosols. This allows for nucleation of droplets around these aerosols, which primarily include dust particles and other, smaller water droplets due to the moisture inside the chamber. While the current experiment acknowledges the presence of such aerosols inside the chamber, it is not possible to control them using the current setup and mechanism. This points to the future scope of more refined cloud chambers, involving the simulation and effects of controlled aerosol injection inside the cloud chamber on the droplet accumulation at different fan speeds within the chamber. Additionally, mechanisms which control the fan speed to introduce turbulence variations inside the cloud chamber during the experiment can also be worked upon, thereby improving the quality of the observations. New ways of measuring the droplet sizes, turbulence source, and some advanced tools for the measurement are already in progress in this field but the variations of the particular parameters and its effect on the atmosphere could provide ground-breaking discoveries.

## References

1. J. D. Spinhirne, and A. E. S. Green, Calculation of the relative influence of cloud layers on received ultraviolet and integrated solar radiation, Atmos. Env. (1967), vol. **12**, no. 12, pp. 2449-2454 (1987) https://doi.org/10.1016/0004-6981(78)90289-5
2. B. A. Wielicki, R. D. Cess, M. D. King, D. A. Randall, and E. F. Harrison, Mission to planet Earth: Role of clouds and radiation in climate, Bulletin of the American Meteorological Society, vol. **76**, no. 11, pp. 2125-2154 (1995) https://doi.org/10.1175/1520-0477(1995)076<2125:MTPERO>2.0.CO;2

3. C. T. R. Wilson, Investigations on lightning discharges and on the electric field of thunderstorms, Proc. R. Soc. Lond. A, vol. **85**, pp. 285-288 (1911) https://doi.org/10.1098/rsta.1921.0003

4. J. L. Katz, Condensation of a supersaturated vapor. I. The homogeneous nucleation of the n-alkanes, The Journal of Chemical Physics, vol. **52**, no. 9, pp. 4733-4748 (1970)

5. H. R. Smith, P. J. Connolly, A. J. Baran, E. Hesse, A. R. D. Smedley, and Ann R. Webb, Cloud chamber laboratory investigations into scattering properties of hollow ice particles, Journal of Quantitative Spectroscopy and Radiative Transfer, vol. **157**, pp. 106-118 (2015) https://doi.org/10.1016/j.jqsrt.2015.02.015

6. M. R. Cholemari & J. H. Arakari, Experiments and a model of turbulent exchange flow in a vertical pipe, International J. Heat and Mass transfer, vol. **48**, pp 4467-4473 (2005) http://dx.doi.org/10.1016/j.ijheatmasstransfer.2005.04.025

7. D. L. Hartmann, M. E. Ockert-Bell, M. L. Michelsen, The effect of cloud type on earth's energy balance: global analysis, Journal of climate, vol. **5**, no. 11, pp. 1281-1304 (1992) https://doi.org/10.1175/15200442(1992)005<1281:TEOCTO>2.0.CO;2

8. T. F. Stocker, D. Qin, GK. Plattner, M. Tignor, S. K. Allen, Judith Boschung, Alexander Nauels, Yu Xia, Vincent Bex, and Pauline M. Midgley, Climate change 2013: The physical science basis, Contribution of working group I to the fifth assessment report of the intergovernmental panel on climate change 1535, (2013)

9. Y. Zhang, A. Macke, and F. Albers, Effect of crystal size spectrum and crystal shape on stratiform cirrus radiative forcing, Atmospheric research, vol. **52**, no. 1-2, pp. 59-75 (1999) https://doi.org/10.1016/S0169-8095(99)00026-5

10. K. Chang, J. Bench, M. Brege, W. Cantrell, K. Chandrakar, D. Ciochetto, C. Mazzoleni, L. R. Mazzoleni, D. Niedermeier, and R. A. Shaw, A laboratory facility to study gas–aerosol–cloud interactions in a turbulent environment: The Π chamber, Bulletin of the American Meteorological Society, vol. **97**, no. 12, pp. 2343-2358 (2016) https://doi.org/10.1175/BAMS-D-15-00203.1

11. R. J. Anderson, R. C. Miller, J. L. Kassner Jr, and D. E. Hagen, A study of homogeneous condensation-freezing nucleation of small water droplets in an expansion cloud chamber, Journal of the Atmospheric Sciences, vol. **37**, no. 11, pp. 2508-2520 (1980) https://doi.org/10.1175/15200469(1980)037<2508:ASOHCF>2.0.CO;2

12. F. Stratmann, A. Kiselev, S. Wurzler, M. Wendisch, J. Heintzenberg, R. J. Charlson, K. Diehl, H. Wex, and S. Schmidt, Laboratory studies and numerical simulations of cloud droplet formation under realistic supersaturation conditions, Journal of Atmospheric and Oceanic Technology, vol. **21**, no. 6, pp. 876-887 (2004) https://doi.org/10.1175/1520-0426(2004)021<0876:LSANSO>2.0.CO;2

13. O. Stetzer, B. Baschek, F. Lüönd, and U. Lohmann, The Zurich Ice Nucleation Chamber (ZINC)-A new instrument to investigate atmospheric ice formation, Aerosol science and technology, vol. **42**, no. 1, pp. 64-74 (2008) https://doi.org/10.1080/02786820701787944

14. J. Lu, H. Nordsiek, E. W.Saw, and R. A. Shaw, Clustering of charged inertial particles in turbulence, Physical review letters, vol. **104**, no. 18, pp. 184505(4) (2010) https://doi.org/10.1103/PhysRevLett.104.184505

15. R. List, J. Hallett, J. Warner, and R. Reinking, The Future of Laboratory Research and Facilities for Cloud Physics and Cloud Chemistry: Report on a Technical Workshop Held in Boulder, Colorado, 20–22 March 1985, Bulletin of the American Meteorological Society, vol. **67**, no. 11, pp. 1389-1397 (1986) https://doi.org/10.2307/26224664

16. P. Rohwetter, J. Kasparian, K. Stelmaszczyk, Z. Hao, S. Henin, N. Lascoux, W. M. Nakaema et al., Laser-induced water condensation in air, Nature Photonics, vol. **4**, no. 7, pp. 451-456, (2010)
17. Y. Petit, S. Henin, J. Kasparian, JP. Wolf, P. Rohwetter, K. Stelmaszczyk, Z. Q. Hao et al., Influence of pulse duration, energy, and focusing on laser-assisted water condensation, Applied Physics Letters, vol. **98**, no. 4, pp. 041105, (2011) https://doi.org/10.1063/1.3546172
18. B. J. Mason, Clouds, rain and rainmaking, Cambridge University Press, ISBN-13: 978-0521157407 (1975)
19. M. Wang, W. Kong, R. Marten et al., Rapid growth of new atmospheric particles by nitric acid and ammonia condensation, Nature, vol. **581**, pp. 184-189, (2020) https://doi.org/10.1038/s41586-020-2270-4
20. S. Thomas, M. Ovchinnikov, F. Yang, D. van der Voort, W. Cantrell, S. K. Krueger, and R. A. Shaw, Scaling of an Atmospheric Model to Simulate Turbulence and Cloud Microphysics in the Pi Chamber, Journal of Advances in Modeling Earth Systems, vol. **11**, no. 7, pp. 1981-1994, (2019) https://doi.org/10.1029/2019MS001670
21. A. Korolev, G. A. Isaac, Relative Humidity in Liquid, Mixed-Phase, and Ice Clouds, Journal of the Atmospheric Sciences, vol. **63**, no. 11, pp. 2865-2880, (2006) https://doi.org/10.1175/JAS3784.1
22. M. S. Mndewa, X. Yuan, D. Huang, B. Li, Effects of light propagation in middle intensity atmospheric turbulence, Front. Optoelectron. China, vol. **2**, no. 3, pp. 312-317, (2009) DOI 10.1007/s12200-009-0030-z
23. W.S. Rasband, ImageJ, U. S. National Institutes of Health, Bethesda, Maryland, USA, https://imagej.nih.gov/ij/, 1997-2018.
24. M.D. Abramoff, P.J. Magalhaes, S.J. Ram, Image Processing with ImageJ, Biophotonics International, vol. **11**, no. 7, pp. 36-42, (2004)
25. Microsoft Corporation, Microsoft Excel, ver. 2016, released July 9, (2015) (MacOS) https://office.microsoft.com
26. K. K. Chandrakar, W. Cantrell, and R. A. Shaw, Influence of Turbulent Fluctuations on Cloud Droplet Size Dispersion and Aerosol Indirect Effects, Journal of the Atmospheric Sciences, vol. **75**, no. 9, pp. 3191-3209, (2018) http://doi.org/10.1175/JAS-D-18-0006.1
27. I. Saito, T. Gotoh, and T. Watanabe, Broadening of Cloud Droplet Size Distributions by Condensation in Turbulence, Journal of the Meteorological Society of Japan, vol. **97**, no. 4, pp. 867-891, (2019) doi:10.2151/jmsj.2019-049
28. Desmos Graphing Calculator, Desmos Graphing Calculator, [online], (2015), Available at: https://www.desmos.com/calculator?create_account